*CORRESPONDENCE
Sachin Maheshwari,
✉ maheshwari.sachin@ed.ac.uk

# Adiabatic capacitive neuron: an energy-efficient functional unit for artificial neural networks

Sachin Maheshwari*, Mike Smart, Himadri Singh Raghav, Themis Prodromakis and Alexander Serb

Institute for Integrated Micro and Nano Systems, University of Edinburgh, Edinburgh, United Kingdom

This paper presents a highly energy-efficient adiabatic capacitive neuron (ACN) hardware implementation of an artificial neuron (AN), with improved energy efficiency, robustness, and scalability over previous work. A single-neuron ACN with 12 one-bit capacitive synapses is implemented in 0.18 µm CMOS technology, supporting both positive and negative synaptic weights. A novel threshold logic (TL) circuit is introduced to realize the binary AN activation function, explicitly designed to minimize input-referred offset and ensure robust decision making under dynamic adiabatic operation. The TL performance is evaluated across three process corners and five temperatures ranging from −55 ˚C to 125 ˚C. Post-layout simulations show that the proposed TL achieves a maximum rising and falling offset voltage of 9 mV, compared to 27 mV (rising) and 5 mV (falling) for a conventional TL implementation across process and temperature variations. The proposed ACN achieves over 90% total synapse energy savings (over 12× improvement) relative to an equivalent non-adiabatic CMOS capacitive neuron (CCN) over operating frequencies from 500 kHz to 100 MHz. A 1000-sample Monte Carlo analysis incorporating process variation and mismatch confirms consistent energy savings exceeding 90% in the synapse energy profile. Supply voltage scaling further demonstrates sustained energy savings above 90%, except for the all-zero input condition, without loss of functionality. These results demonstrate that adiabatic charge recovery, combined with a robust low-offset threshold logic design, enables substantial energy reduction while maintaining reliable neuron operation across wide operating conditions.

## 1 Introduction

Adiabatic Logic (AL) is a charge recovery design technique that operates with a gradually alternating AC power supply that periodically returns capacitive charge to the supply (Teichmann, 2012). This significantly differs from traditional, non-adiabatic, CMOS solutions that use a fixed DC supply. Appreciating the complexities of designing an energy-efficient AC power supply (Raghav et al., 2017; Jeanniot et al., 2017; Raghav et al., 2016; Maksimovic et al., 2000), previous AL work has successfully demonstrated the potential for significant energy savings (Blotti and Saletti, 2004; Oklobdzija et al., 1997; Raghav and Bartlett, 2020; Tang and Liter, 2010; Maheshwari and Kale, 2019; Houri et al., 2015; Maheshwari et al., 2018). It has been actively researched, including the novel implementation of an adiabatic hardware description language model, which shows timing agreement with transistor-level simulations (Maheshwari et al., 2021a).

Additionally, it has been integrated with other low-power techniques and emerging devices (Kahleifeh et al., 2023), as well as an adiabatic leaky integrate and fire neuron featuring a refractory period and ultra-low energy operation Massarotto et al. (2024). As such, AL techniques are ideal for the implementation of power-hungry Artificial Neural Networks (ANNs). This includes capacitive ANN solutions (Maheshwari et al., 2021b; Ozdemir et al., 1996; Padure et al., 1999) and, most recently, with emerging memcapacitor configurable devices (Demasius et al., 2021). The simplest mathematical model of an artificial neuron is comprised of a vector dot-product between a number of input signals and a corresponding set of weights (or *synapses*) Botros and Abdul-Aziz (1993) followed by a non-linear activation function Lippmann (1987). The neuron output can then act as an input to other ANs in a multi-layer, multi-AN network. In digital hardware, this translates to large numbers of power-hungry multiply-accumulate operations. Optimizing artificial neurons and synapses for energy is, therefore, a key target for building efficient artificial intelligence systems.

Substantial work has been done on the implementation of synapses in hardware using devices such as resistors, MOSFETs, sink or source DC currents, and capacitors (Teng, 2006; Çilingiroğlu, 1991; Verleysen et al., 1989; Pelayo et al., 2005). Out of these, capacitive solutions are the most desirable because of their flexibility in fabrication technology, simple sensing, reduced sensitivity to process variations compared to active devices and high energy efficiency. Over the years, many Switched-Capacitor (SC) neural networks have been proposed to perform computations, such as the analog dot product, instead of using traditional digital methods (Bankman and Murmann, 2017; Tang et al., 2022). The earliest implementation demonstrated energy efficiency at the expense of increased complexity (Tsividis and Anastassiou, 1987). Another work presented a fully synchronous SC-based self-organizing analog neural network with a winner-take-all circuit, capable of computing the AN dot product (Maundy and El-Masry, 1991). Further SC implementations include a differential comparator-based charge redistribution design with symmetric capacitor banks on either side (López-García et al., 2004). This configuration offers better stability but dissipates energy by transferring charge from the supply to ground during each reset. Capacitor leakage over time further degrades functionality and increases energy consumption. Using a sinusoidal AC supply with energy recovery via charge transfer could potentially mitigate leakage losses.

In parallel with SC and AL developments, there has also been significant recent research into Binary Neural Networks (BNNs) that use neurons with binary inputs and activation functions (Qin et al., 2020). Work in this area has been driven by the desire for fast and low-resource (memory) AN implementations in digital hardware (Yayla et al., 2023). Consequently, BNNs are trained with binary, ternary, or heavily quantized, positive and negative-valued weights for optimal storage and efficient computation. Researchers have shown that BNNs with more neurons can match the classification performance of state-of-the-art ReLU-based ANNs on datasets like MNIST and CIFAR-10 (Hubara et al., 2018). Novel AL-based SC hardware with binary I/O neurons is therefore well-suited for BNNs.

The primary goal of this work is to introduce a complete and highly energy-efficient, analog, fixed (or *baked-in*) hardware implementation of an, well-suited for energy-intensive feature extraction ANN front-ends, such as those used in transfer learning (Weiss et al., 2016). As a secondary aim, we refine and extend the Adiabatic Capacitive Artificial Neuron (ACAN) architecture presented in the authors' prior work (Maheshwari et al., 2022). While the earlier study provided a detailed and practical analysis, it did not consider the accurate circuit-level mapping of real-world artificial neurons. In this next-generation design, as illustrated in Figure 1, functional support for both real-valued positive and negative weights is introduced, susceptibility to variations in the adiabatic power clock is significantly reduced and the accuracy of the binary threshold logic is improved via a new low-offset TL circuit. Nonetheless, prior work on switched-capacitor (SC) neural networks—starting with the seminal work of (Tsividis and Anastassiou, 1987)—relied on proportional mapping schemes to convert artificial-neuron weights into capacitance values. Later efforts examined dot-product computation in non-adiabatic SC networks by explicitly relating stored charge to synaptic weights and capacitor sizing Maundy and El-Masry (1991). In contrast, the present ACN approach enables energy-efficient adiabatic operation while preserving accurate neuron-level mapping. Finally, the paper discusses the benefits of the highly scalable properties of the new design and compares it with a purely CMOS-based equivalent solution. Through comparative analysis and post-layout based validation, this work aims to demonstrate how the adiabatic technique can enhance energy efficiency by robustly mapping real-valued AN weights to ACN circuits to generate equivalent functionality.

## 2 Adiabatic capacitive neuron: Design overview

An artificial neuron comprises a set of weights (*synapses*) and an activation function. This paper considers an artificial neuron with a Heaviside activation function that has $N$ binary inputs, $x_i$ where $i$ is an element of the indexing set $I = \{0.N-1\}$, and has a single binary output, $y$. The output is expressed as

$$y = \begin{cases} 1, \text{ if } \sum_{i \in I} w_i x_i \geq \tau \\ 0, \text{ otherwise} \end{cases} \tag{1}$$

where $w_i$ are $N$ trained weight values and $\tau$ is a constant bias value. The weights $N$ may be real-valued or quantized and can be split into two disjoint sets of $N^+$ positive-valued (*excitatory*) weights, $w_i^+$ where $i \in I^+$, and $N^-$ negative-valued (*inhibitory*) weights, $w_i^-$ where $i \in I^-$. $I^\pm$ are disjoint indexing subsets of $I$ such that $N = N^+ + N^-$.

Figure 1 introduces a Double-Tree Single-Clock (DTSC) implementation of an ACN comprising two capacitive trees and a single sinusoidal Power Clock (PC). It includes a minimal set of $N$ synapse capacitors required to embody the AN weights. This includes a subset of $N^+$ synapse capacitors, $C_i^+$, in the first (positive) capacitive tree. The capacitance values of each $C_i^+$ map from the set of $N^+$ positive-valued AN weights, $w_i^+$, defined in Equation 1. It also has $N^-$ synapse capacitors, $C_i^-$, in the second (negative) capacitive tree with capacitance values mapped from the magnitude of each negative-valued AN weight, $w_i^-$.

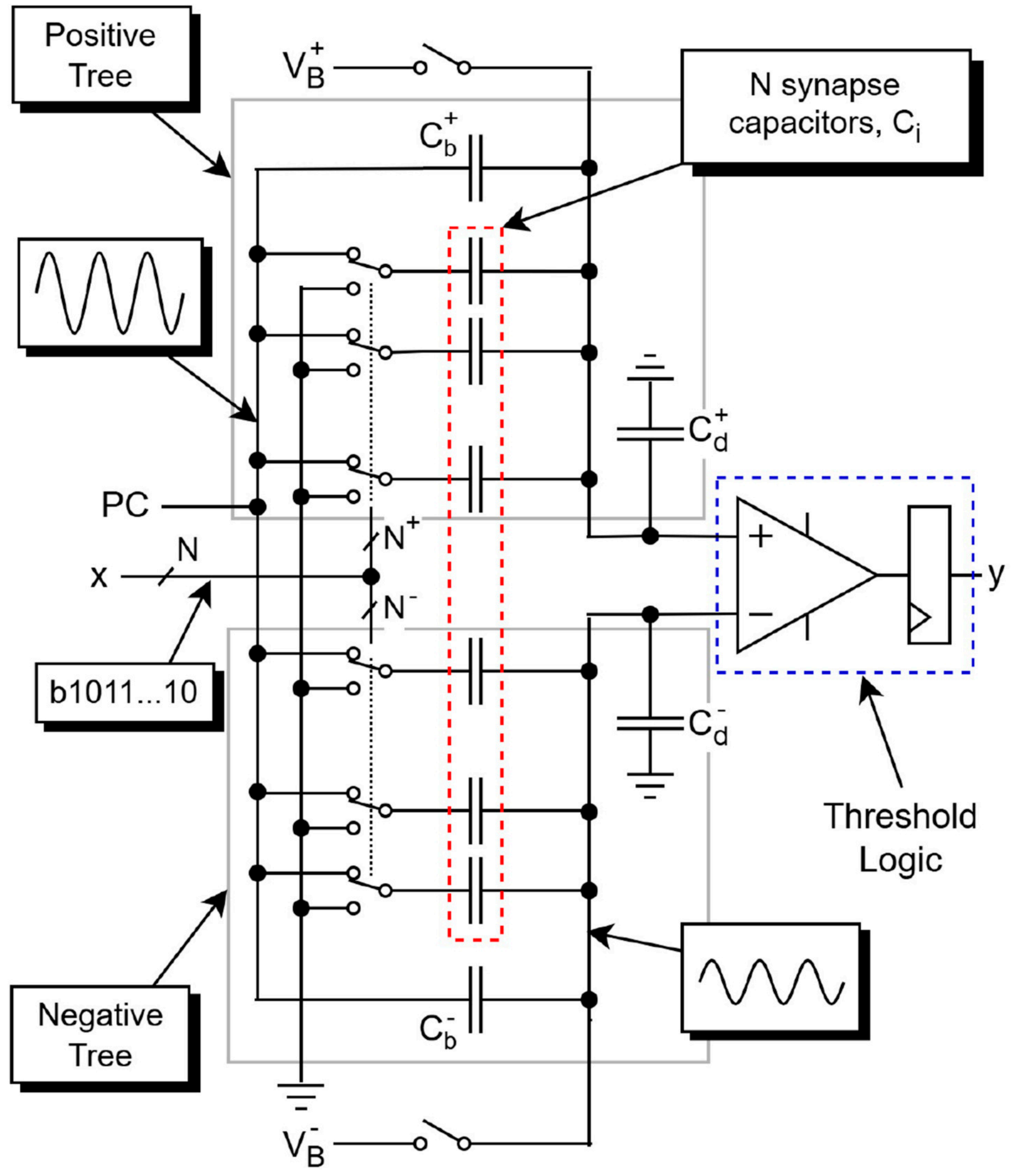

FIGURE 1
N-input Double-Tree Single-Clock (DTSC) ACN design. The design consists of two sections: a capacitive tree with switches and the threshold logic. The capacitive tree comprises positive and negative branches, each having synapse, bias and ballast capacitors with SPDT switches. Each branches also have a bias voltage, $V_B \pm$ to bias the membrane node, $v_m$, externally. The threshold logic is implemented using a Dynamic Latch Clocked Comparator (DLCC) and an SR latch.

A set of $N$ Single-Pole Double-Throw (SPDT) switches is associated with each of the $N$ synapse capacitors in the two capacitive trees. Each switch acts as a charge-steering element, synchronized with the slowly varying power-clock that connects to a synapse capacitor $C_i$ bottom plate. When the switch connected to $x_i$ is active, the rising and falling phases of the PC cause the corresponding node voltages to track the supply ramps smoothly, enabling near-adiabatic energy transfer. If the switch is inactive, then the corresponding $C_i$ bottom plate is grounded. The modulation of the inputs combined with their synapse capacitances generates two sinusoidal *membrane voltages*, $v_m^{\pm}$, at the input terminals of the threshold logic. The TL, which implements the AN activation function, then generates the final output, $y$, based on the two input membrane voltage values. If $v_m^+ > v_m^-$ at the time of sampling, the comparator outputs a binary value of 1, otherwise, 0. The result is a hardware implementation of AN defined in Equation 1, with the switched capacitors allowing for energy-efficient, adiabatic operation.

The ACN also includes bias capacitors ($C_b^{\pm}$) to support the software bias term, $\tau$, ballast capacitors ($C_d^{\pm}$) and DC bias voltages ($V_B^{\pm}$) connected via a Transmission Gate (TG) switch to the positive and negative terminals of the comparator. The $C_b^{\pm}$ and $C_d^{\pm}$ capacitors are important as they control the swing amplitude of $v_m^{\pm}$ at the comparator inputs. The maximum number of necessary capacitors in an N-input ACN is $N + 4$. Note, under certain conditions, the bias and/or ballast capacitors can be omitted.

## 2.1 Adiabatic capacitive tree network

A single SPDT-capacitive synapse with a TG reset/bias switch for $v_m$, along with bias and ballast capacitors, is depicted in Figure 2a. Figure 2b shows the transistor-level diagram of a single capacitive synapse with a ballast capacitor; $V_B$ is omitted for clarity. The PC is instrumental in the working of the capacitive tree. The time-varying sinusoidal PC signal voltage, $V_{pc}(t)$, varies from rail-to-rail, to enable computation and charge recovery. It operates in two modes, namely,: $Reset\ Mode$ and $Operational\ Mode$. In $Reset\ Mode$, the system is in an idle state, where the PC is resting at its minimum level. The $Operational\ Mode$ features a sinusoidal, wave-like behaviour and is divided into two phases. During the upswing (rising) of each PC voltage wave, the system is in the $Evaluation\ Phase$ and charge enters through the SPDT switches onto the synapse capacitors. Conversely, during the downswing (falling) of the PC wave, the system enters the $Recovery\ Phase$ and charge recedes from the synapse capacitors via SPDT switches back to the specially designed Power Clock Generator (PCG) (Maheshwari et al., 2022), allowing near-adiabatic energy exchange. The different modes and phases are illustrated in Figure 2c. The parasitic capacitances associated with SPDT synapse switches primarily impact overall energy consumption through additional charge and discharge events, while having a negligible effect on the $v_m^{\pm}$ node voltage due to their isolation from the charge-integration path (Maheshwari et al., 2022).

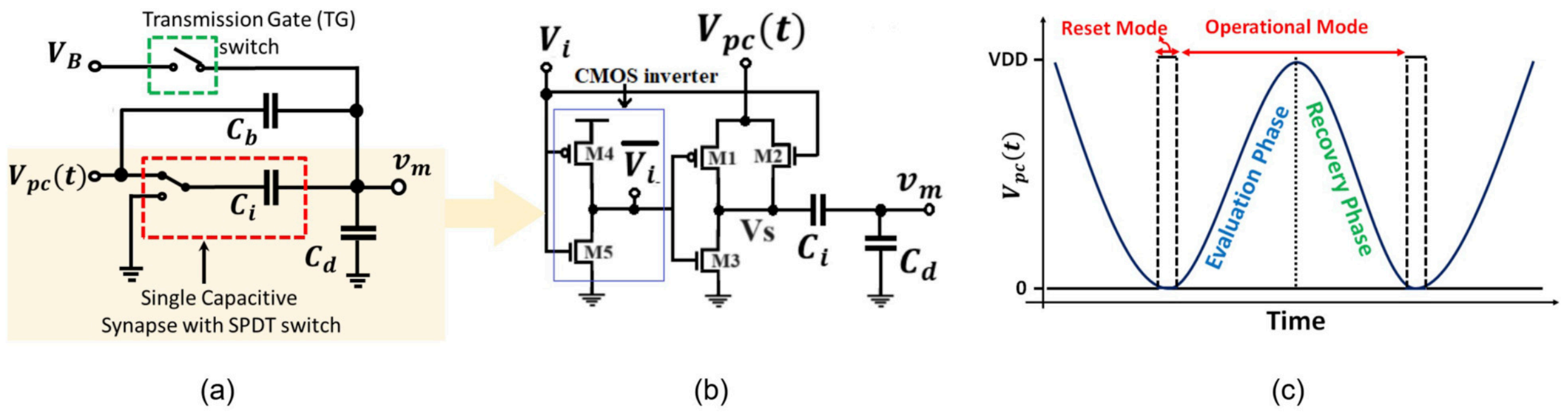

FIGURE 2
**(a)** A single capacitive SPDT synapse switch showing synapse capacitor, $C_i$, bias capacitor, $C_b$, ballast capacitor, $C_d$, and the bias voltage, $V_B$. The MIMCaps are used as capacitors from the 180*nm* Process Design Kit (PDK). **(b)** Transistor-level diagram for a single capacitive SPDT synapse switch with synapse and ballast capacitances. The inverted signal, $\overline{V_i}$, is generated using an inverter. **(c)** Power clock sinusoidal voltage wave showing the two modes: *reset* and *operational* modes. The circuit works in the operational mode, which is divided into *Evaluation* and *Recovery* Phase.

The single-pole single-throw, TG switch in Figure 2a is connected to a constant bias voltage, $V_B$, and is on (closed) in *Reset Mode*. The membrane voltage $v_m$ consequently takes on the value of $V_B$ at this time, providing a stable initial, minimum $v_m$ *throughout the entire cycle*. When computation begins, the TG switch opens and $v_m$ depends on the state of the synapse switch and capacitance. When the input is zero ($x_i = 0$), the switch is connected to ground via transistor $M3$ (see Figure 2b) and appears parallel to the ballast capacitor, effectively summing together as $C_i + C_d$ at $v_m$. When the input, $x_i = 1$, the synapse capacitor is connected to the PC via transistors $M1$ and $M2$. During *Evaluation Phase*, the synapse and bias capacitor forming a divider with $C_d$ and parasitic capacitance to ground at the $v_m$ node starts charging. The parasitic capacitances at the $v_m$ node originate primarily from the top plate parasitic capacitance of $C_i$ and the input transistors of the TL stage. To simplify the analysis and maintain clarity in the modeling, these parasitic contributions are explicitly lumped post-mapping into a single ballast capacitance term, $C_d$, in the equations presented below.

Considering the general case, with $N$ synapses distributed across two capacitive trees, the comparator membrane voltages, $v_m^{\pm}$, at time $t$, can be determined by standard capacitive voltage division as

$$v_m^{\pm}(t) = V_B^{\pm} + V_{pc}(t)\left[\sum_{i\in I^{\pm}} \frac{C_i^{\pm} x_i}{C_A^{\pm}} + \frac{C_b^{\pm}}{C_A^{\pm}}\right] \tag{2}$$

where the denominator terms $C_A^{\pm} = C_T^{\pm} + C_b^{\pm} + C_d^{\pm}$ represents the total capacitance in each tree and $C_T^{\pm} = \sum_{i\in I^{\pm}} C_i^{\pm}$ are the total synapse capacitances per tree. The mathematical proof for the above equation is elaborately given in (Smart et al., 2025). The use of SPDT synapse switches, compared to the TG synapse switches used in ACAN, means that $C_A^{\pm}$ is constant and independent of the input $x_i$. There is now a linear relationship between $v_m$ and $x_i$, similar to that defined by the software AN condition in Equation 1. The membrane voltages can also be expressed as shown in Equation 3

$$v_m^{\pm}(t) = V_B^{\pm} + V_{pc}(t)\frac{C_{on}^{\pm}}{C_{on}^{\pm} + C_{off}^{\pm}} \tag{3}$$

where $C_{on}^{\pm}$ is the sum of the switched on capacitor values ($x_i = 1$) plus the bias capacitor $C_b^{\pm}$, and $C_{off}^{\pm}$ is the sum of all the switched off capacitors ($x_i = 0$) connected to ground, plus the ballast capacitor, $C_d^{\pm}$.

A proportional mapping scheme can be applied to convert the abstract software AN weights to physical ACN capacitance values, as originally proposed in (Tsividis and Anastassiou, 1987), for functional equivalence, as defined by

$$C_i^{\pm} = \frac{|w_i| C_T}{w_T} \tag{4}$$

where $w_T = \sum_i |w_i|$ and $C_T = C_T^+ + C_T^-$. The value of $C_T$ can be considered a design choice that controls the total physical area of the ACN. The software bias, $\tau$, can be mapped in the same way onto one of the bias capacitors, depending on the sign. It should be noted that there is a requirement to train the AN weights and select a value of $C_T$ such that the minimum $C_i$ is greater than, or equal to, the minimum supported by the technology, $C_{min}$. The design choice of $C_T$, and properties related to the mapping strategies are presented in (Smart et al., 2025).

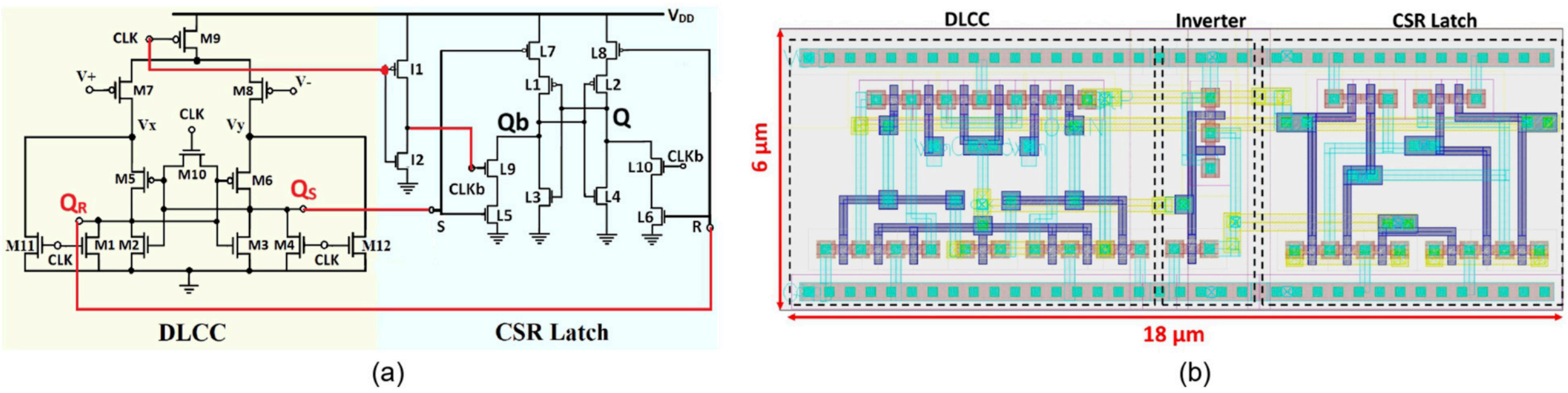

FIGURE 3
**(a)** Transistor-level diagram of the threshold logic showing two stages. The first stage is a pMOS-based DLCC (yellow shade) and the second stage is a proposed clocked SR latch (blue shade). The complementary clock *CLKb*, for the SR latch is derived on-chip by inverting the external *CLK* signal. **(b)** The layout of the proposed TL circuit comprises a two-stage architecture with an inverter that generates a complementary clock for the latch. To reduce mismatch-induced offset and improve robustness, the layout is arranged as symmetrically as possible, with matched device placement and balanced routing, while all transistors are implemented using minimum feature sizes to achieve a compact footprint.

The fixed capacitor values control the swings of the membrane voltages during each PC waveform, $C_i^{\pm}$, representing the AN weights, and the inputs, $x_i$. As such, the extent of the swing range at the peak of the wave lies between the two conditions when all inputs are logic '0' and when all inputs are logic '1', as expressed in Equation 5

$$v_m^{\pm}(t) = \begin{cases} V_B^{\pm} + V_{pc}(t)C_b^{\pm} \big/ C_A^{\pm}, \forall x_i = 0 \\ V_B^{\pm} + V_{pc}(t)\left(C_T^{\pm} + C_b^{\pm}\right) \big/ C_A^{\pm}, \forall x_i = 1 \end{cases} \tag{5}$$

On the other hand, ACAN (Maheshwari et al., 2022) requires a suitably large-valued ballast capacitor as a mandatory requirement. However, in ACN, some ballast capacitance is naturally provided from SPDT-grounded synapse capacitors. The two ballast capacitors, $C_d^{\pm}$, are typically still required and, importantly, act as asymmetric scaling terms to balance the two capacitive trees. This novel balancing functionality allows for the mapping of $N$ software weights to a minimal set of $N$ capacitors, whilst providing identical functionality compared to the software AN, as defined by Equation 1. The derivation of the specific capacitance values of $C_b^{\pm}$ and $C_d^{\pm}$, which are required to complete this mapping, is discussed in detail in (Smart et al., 2025).

## 2.2 Threshold logic

The Threshold Logic (TL) unit in the DTSC ACN implements the neuron's activation function. The two membrane voltages generated by the capacitive tree network in Section 2.1 serve as inputs to the TL circuitry. The improved TL design, introduced in this paper, includes two stages as depicted in Figure 3. The first stage is a Dynamic Latch Clocked Comparator (DLCC), and the second stage is a proposed Clocked Set-Reset (CSR) latch. A pMOS variant is preferred over an nMOS variant, as it eliminates the need for external biasing to keep the membrane potential above the subthreshold region, thereby reducing energy consumption. The complementary clock $CLKb$, for the proposed clocked SR latch, is derived internally by inverting the external $CLK$ signal. More conventional designs, such as the one from (Matsui et al., 1994), which we use to benchmark against, use a NAND/NOR-based SR instead of the CSR latch (see Supplementary Figure S1). Conventional TL design suffers from two major issues (Matsui et al., 1994), namely, 1) large propagation delay, affecting the offset voltage, and 2) latching the wrong data with reduced voltage headroom.

TABLE 1 **Post-layout rising offset voltage (*mV*) of the conventional (*Conv*) and the proposed (*Prop*) TL design across process corners, and temperatures.**

| Temperature | FF | | TT | | SS | |
|---|---|---|---|---|---|---|
| °C | Conv | Prop | Conv | Prop | Conv | Prop |
| −55 | 23.00 | 9.003 | 17.00 | 7.005 | 13.01 | 5.007 |
| 0 | 23.00 | 9.004 | 19.00 | 9.005 | 15.01 | 7.007 |
| 27 | 25.00 | 9.004 | 21.00 | 9.005 | 15.01 | 7.007 |
| 100 | 27.00 | 9.004 | 23.01 | 9.006 | 19.01 | 9.008 |
| 125 | 27.00 | 9.004 | 23.01 | 9.006 | 19.01 | 9.008 |

With reference to Figure 3a, during the pre-charge phase, the clock signal ($CLK$) transits from zero to $V_{DD}$. Consequently, $M9$ is turned off and $M1$, $M4$, $M11$ and $M12$ are turned on. As a consequence, the output nodes ($Q_R$ and $Q_S$) of the DLCC are pre-charged to zero and the DC path from the supply to the ground is cut off. As such, a second stage is necessary to latch the output correctly. During the pre-charge phase, the second stage $CLKb$ signal transits from $V_{DD}$ to zero and the transistors $L9$ & $L10$ and $L5$ & $L6$ are switched off, thus the CSR latch holds the previous state of $Q$ and $Qb$, giving a stable output for each clock period.

Next, in the comparison phase, the $CLK$ signal transitions from $V_{DD}$ to zero, and the transistors $M1$, $M4$, $M11$, and $M12$ are turned off, while the $M9$ transistor is switched on. The DLCC starts by comparing the two input voltages: $v_m^-$ and $v_m^+$, resulting in the output to swing differentially, causing one output node to move to $V_{DD}$, and the other complementary output node to ground. Assuming $v_m^+ > v_m^-$, the output $Q_S$ is pulled to $V_{DD}$ and due to the positive feedback transistors, $Q_R$ is pulled down to $0V$. The output from the first stage is fed as input to the CSR latch. Here, $Q_S$ is connected to the set $S$ terminal and $Q_R$ to the reset $R$ terminal. As a result, transistor $L6$ starts conducting, whereas $L5$ is switched off. Since the transistors $L9$ & $L10$ are already on, the output node $Qb$ discharges to ground and due to the positive feedback, the node $Q$ charges to $V_{DD}$. The layout plays an important role in the offset voltage and resolution. To keep the TL offset as low as possible, the layout is drawn symmetrically and is shown in Figure 3b. The Supplementary Section 1 reports a comparison between the conventional and proposed TL, and the waveform is shown in Supplementary Figure S2.

TABLE 2 Post-layout falling offset voltage ($mV$) of the *Conv* and (*Prop*) TL design across process corners, and temperatures.

| Temperature | FF | | TT | | SS | |
|---|---|---|---|---|---|---|
| °C | Conv | Prop | Conv | Prop | Conv | Prop |
| −55 | 2.997 | 8.996 | 0.9956 | 4.995 | −1.006 | 2.993 |
| 0 | 4.996 | 8.996 | 2.995 | 6.995 | 0.9927 | 4.993 |
| 27 | 4.996 | 8.996 | 2.995 | 6.995 | 0.9928 | 4.993 |
| 100 | 4.996 | 8.995 | 2.995 | 8.994 | 0.9927 | 6.992 |
| 125 | 4.996 | 6.996 | 2.995 | 8.994 | 0.9926 | 6.992 |

The post-layout rising/falling offsets and energy consumption are evaluated across three process corners and temperatures from $-55\ ^oC$ to $125\ ^oC$. The testbench setup, shown in Supplementary Figure S3, with a $CLK$ frequency of $1MHz$ and an input voltage range $V_{in}$ of $-0.1V$ to $+0.1V$ with a common mode voltage $V_{cm}$ = $0.8V$, giving the input resolution of $1mV/\mu s$. Based on Tables 1, 2, the conventional TL shows a large asymmetry in the offsets. It has a large rising post-layout offset of 10 s of $mV$, ranging from $27mV$ at 125 $^oC$ FF corner to $13mV$ at $-55\ ^oC$ SS corner, while the proposed TL has a range from $9mV$ to $5mV$ for the same corners and temperature. On the other hand, the conventional TL shows a lower falling offset ranging from $5mV$ at 125 $^oC$ FF corner to $-1mV$ at $-55\ ^oC$ SS corner, while the proposed TL ranges from $3mV$ to $9mV$. Overall, the proposed TL shows greater offset symmetry and consistency. On the other hand, the simulated energy across corners and temperatures is shown in Supplementary Figure S4, with the proposed TL reducing the average energy by 1.5% (SS) and 2.3% (FF) vs. the conventional design.

Finally, the TL typically samples the membrane voltages at the peak of the PC clock when $V_{pc}(t) = V_{max}$, at the end of *Evaluation Phase*. Combining this information with Equation 2, the TL will produce output-high under the condition shown in Equation 6 and output-low otherwise:

$$V_B^+ + V_{max}\sum_{i\in I^+}\frac{C_i^+ x_i + C_b^+}{C_A^+} \geq V_B^- + V_{max}\sum_{i\in I^-}\frac{C_i^- x_i + C_b^-}{C_A^-} \quad (6)$$

As discussed in the previous section, the ballast capacitance values, $C_d^\pm$, in $C_A^\pm$ can be used to scale one or both sides of the condition in Equation 6. This scaling can be used to ensure that the swing range of the membrane voltages $v_m^\pm$ is always within the operational voltage range of the comparator $[0,\ V_{cut}]$, where $V_{cut}$ may be significantly lower than $V_{max}$. $V_{cut}$ is approximately $V_{DD} - |V_{thp}|$, where $V_{thp}$ is the threshold voltage of the pMOS transistors used in the comparator. Importantly, owing to the differential design of the ACN, the sampling timing does not need to be highly precise. While sampling ideally should occur at the peak, $v_m^{peak}$, the system can tolerate reasonable timing skew with only a modest reduction in $v_m^{peak}$, with potential classification errors arising only under extreme timing mismatches. A Supplementary Figure S5 demonstrates that our DTSC network can tolerate dynamic sampling of the differential voltages, $v_m^\pm$, thus meeting timing closure of the TL. The post-layout transient simulations are shown for two representative test vectors, TV4 and TV12, selected to capture worst-case loading and near-threshold decision conditions. Additionally, the proposed architecture is not restricted to real-valued weights and can readily accommodate quantized as well as binary synaptic weights (+1/−1) (Smart et al., 2025; Yayla et al., 2023).

TABLE 3 DTSC $N = 12$ ACN configuration with $N^+ = 5$, $I^+ = \{0,5,6,9,10\}$ and $N^- = 7$, $I^- = \{1,2,3,4,7,8,11\}$ where $i$ is the synapse index, $w_i$ the abstract weight and $C_i$ the corresponding synaptic capacitance.

| $i$ | $w_i$ | $C_i$ $(fF)$ |
|---|---|---|
| 0 | 0.937 | 195 |
| 1 | −1.000 | 208 |
| 2 | −1.000 | 208 |
| 3 | −1.000 | 208 |
| 4 | −1.000 | 208 |
| 5 | 0.169 | 35 |
| 6 | 0.600 | 125 |
| 7 | −1.000 | 208 |
| 8 | −0.529 | 110 |
| 9 | 0.992 | 206 |
| 10 | 0.961 | 200 |
| 11 | −1.000 | 208 |
| $C_b^+$ $(fF)$ | 35 | |
| $C_b^-$ $(fF)$ | 56 | |
| $C_d^+$ $(fF)$ | 1159 | |
| $C_d^-$ $(fF)$ | 543 | |

# 3 Hardware implementation of ACN

The hardware implementation of a single neuron with 12 1-bit input synapses, ACN, is done in Cadence EDA tool using a $0.18\mu m$ commercially available CMOS technology at a nominal $V_{DD} = 1.8\ V$ supply and $1MHz$ operating frequency. The PC is generated using the PCG discussed in Supplementary Section 3, while a detailed analysis is provided in (Maheshwari et al., 2022; Raghav et al., 2025). All SPDT synapse switches are kept at a technology minimum width. Synapse capacitors are implemented using MIMCAPs, whose values are chosen based on the mapping Equation 4, and quantized to allowable MIMCAP widths and lengths. Bias and ballast capacitance values are selected using the optimal *conditional mapping* process defined in Smart et al. (2025). Table 3 shows the mapped configuration for a single ACN, generated using weights extracted from a randomly selected software AN with a fixed $\tau = 0.1$, that includes a selection of both positive and negative weights as well as varying weight magnitudes. The ACN configuration uses: $V_{max} = 1.8V$, $V_B^{\pm} = 0V$, $C_{min} = 35fF$ and $V_{cut} = 1.3V$. The total capacitance of this ACN is $3907fF$ with a total synapse capacitance, $C_T = 2115fF$.

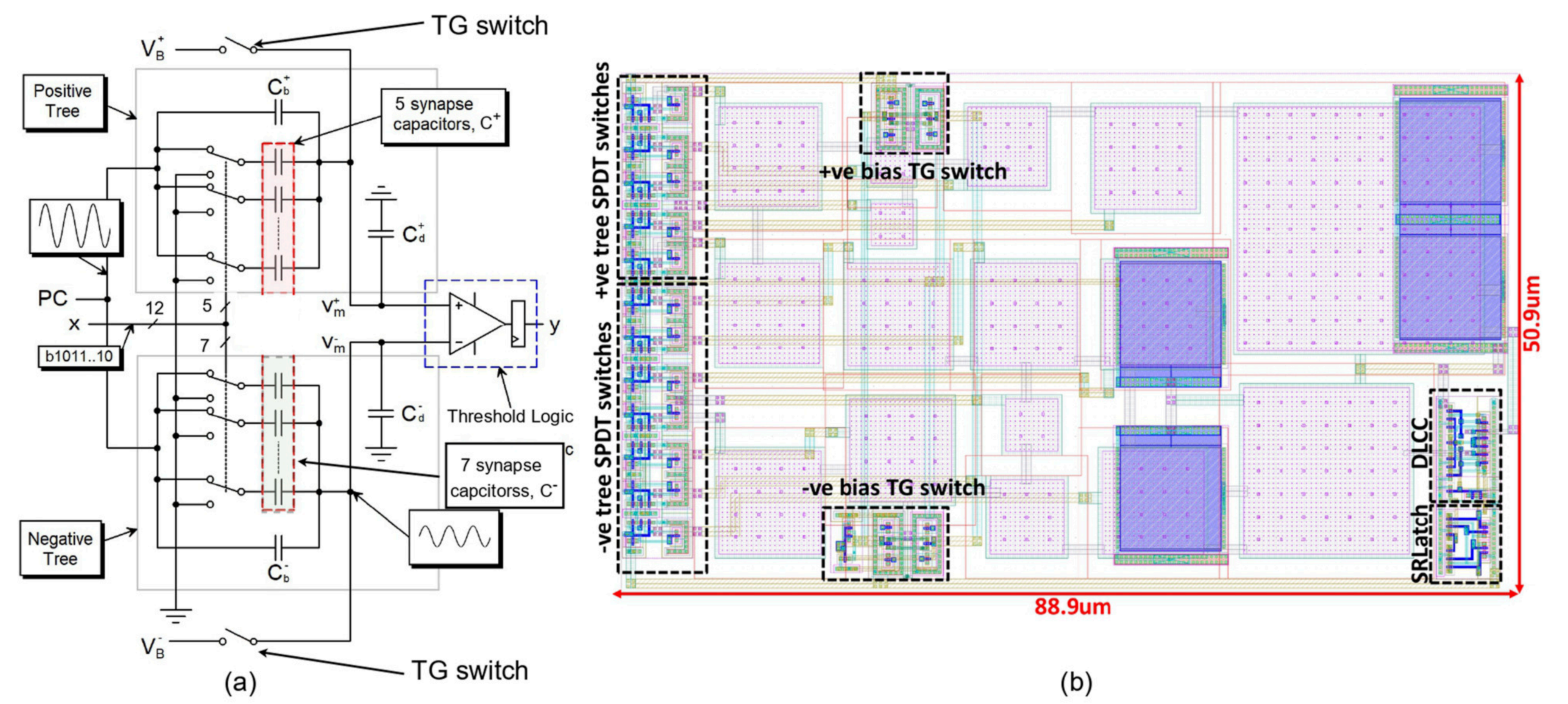

FIGURE 4
**(a)** 12-input Double-Tree Single-Clock (DTSC) ACN design with 5 positive (top red rectangle) and 7 negative (bottom green) synapse capacitor trees. The *PC* is given from the PCG, shown in the Supplementary Figure S7. **(b)** The ACN Layout in **0.18$\mu$m** CMOS technology. To ensure layout uniformity, the positive and negative synapse SPDT switch devices are placed on the left side of the layout, while the bias switches are placed above the MIM capacitor array. The TL block is positioned on the right, clearly defining the input and output routing interfaces. The dark-shaded blue regions denote reversed diodes included to mitigate antenna effects during fabrication.

Using the parameters defined in Table 3, the capacitive synapse trees are instantiated in the design and integrated with the TL gate to complete the computation based on the input signals, $x_i$. Although not optimal, both stages of the TL were designed using minimum-size transistors for minimum energy consumption. The physical layout of the circuit with RC extraction was performed before simulating the circuit for hardware and software comparison. The post-layout extracted result shows an extra parasitic capacitance of approximately $30fF$ value on each of the $v_m^{\pm}$ nodes, acting parallel to the $C_d^{\pm}$. Therefore, these parasitic values can be compensated for within the ballast capacitance values during fine-tuning. The ACN circuit diagram and its layout are shown in Figure 4. Each capacitor, having equal length and width, is oriented in an array format for uniform connectivity and area. The SPDT synapse switches are on the extreme left side of the layout, which will be connected to the 12 1-bit inputs and the PC. The TL gate is on the right side of the layout, next to the negative ballast capacitor. Large antenna diodes (reverse diodes) having $W/L = 10\mu\text{m}/10\mu\text{m}$ were used to reduce antenna effects during fabrication. The unoptimized dimension for our single neuron with 12 1-bit input synapses is **88.9$\mu$m × 50.9$\mu$m**.

# 4 Results

To analyse ACN performance, a subset of test vectors from the original weight-training dataset was selected, along with a number of input corner cases, and is reported in the Supplementary Table S1. In total, 16 test vectors were chosen based on two factors: 1) predicted input differential voltage ($v_{md} = v_m^+ - v_m^-$), to test the accuracy of the TL offset, and 2) minimum to maximum ACN energy dissipation levels based on the computed capacitive load of the PC. See Supplementary Section 4 for the derivation of the capacitive load.

Performance evaluation is based on the full custom post-layout for all the designs using the Spectre simulator. The PCG circuit parameters defined in a previous paper Maheshwari et al. (2022) are used in this work and are set as follows: bypass-switch on period, $t_{ON} = 60\ ns$; power supply capacitance, $C_E = 25\ pF$ and inductance, $L_{PC} = 1\ mH$, that generates a nominal power clock frequency of $1\ MHz$. An additional ideal capacitive load of $100\ fF$ is attached to the output of the TL gate throughout the post-layout simulation. The working and generation of the *PC* signal can be found in the Supplementary Section 3.

## 4.1 Functionality

To verify functionality, the ACN output obtained with the improved TL design is compared with that of the conventional TL-based ACN and the theoretical adiabatic model. For each input test vector, the theoretical model predicts the peak membrane voltages, when $V_{pc}(t) = V_{max}$, based on Equation 2, and the TL output based on condition in Equation 6. Table 4 compares the theoretical model against Cadence post-layout hardware simulations. It is also worth reporting that the theoretical ACN model output defined by Equation 2 has been verified to match that of the software ANN as defined in Equation 1.

TABLE 4 Comparison between the theoretical model, proposed ACN and ACN using conventional TL design. The red highlight box in the table indicates discrepancies between theoretical and circuit outputs.

| | Theoretical model | | | | Proposed ACN | | | Conventional ACN | | |
|---|---|---|---|---|---|---|---|---|---|---|
| Test Vectors | $v_m^+$ (mV) | $v_m^-$ (mV) | $v_{md}$ (mV) | Output | $v_m^+$ (mV) | $v_m^-$ (mV) | Output | $v_m^+$ (mV) | $v_m^-$ (mV) | Output |
| TV1 | 32.0 | 1,301.0 | −1,268.0 | 0 | 34.8 | 1,258.4 | 0 | 34.6 | 1,257.1 | 0 |
| TV2 | 733.0 | 1,301.0 | −568.0 | 0 | 707.8 | 1,261.0 | 0 | 705.8 | 1,257.4 | 0 |
| TV3 | 147.0 | 434.0 | −287.0 | 0 | 145.0 | 426.3 | 0 | 146.4 | 427.3 | 0 |
| TV4 | 733.0 | 918.0 | −185.0 | 0 | 700.2 | 875.9 | 0 | 699.6 | 875.1 | 0 |
| TV5 | 32.0 | 153.0 | −120.0 | 0 | 32.2 | 149.1 | 0 | 32.1 | 149.1 | 0 |
| TV6 | 549.0 | 625.0 | −77.0 | 0 | 534.4 | 607.3 | 0 | 533.8 | 606.6 | 0 |
| TV7 | 701.0 | 727.0 | −26.0 | 0 | 674.2 | 699.5 | 0 | 672.4 | 697.6 | 0 |
| TV8 | 32.2 | 51.5 | −19.3 | 0 | 31.3 | 49.4 | 0 | 31.0 | 49.2 | 0 |
| TV9 | 147.0 | 153.0 | −5.0 | 0 | 143.3 | 149.2 | 0 | 143.4 | 149.2 | 0 |
| TV10 | 244.0 | 243.0 | 1.0 | 1 | 239.2 | 238.4 | 0 | 239.1 | 238.3 | 0 |
| TV11 | 733.0 | 727.0 | 6.0 | 1 | 707.6 | 701.3 | 1 | 707.3 | 700.8 | 0 |
| TV12 | 553.0 | 535.0 | 18.0 | 1 | 537.8 | 520.5 | 1 | 537.4 | 520.6 | 0 |
| TV13 | 586.0 | 535.0 | 50.0 | 1 | 573.4 | 525.6 | 1 | 574.6 | 526.5 | 1 |
| TV14 | 359.0 | 243.0 | 116.0 | 1 | 353.6 | 240.2 | 1 | 353.6 | 240.2 | 1 |
| TV15 | 327 | 52.0 | 275.0 | 1 | 316.7 | 52.3 | 1 | 317.5 | 52.2 | 1 |
| TV16 | 733.0 | 52.0 | 681 | 1 | 716.5 | 55.5 | 1 | 715.9 | 55.3 | 1 |

TABLE 5 The comparison of the total synapse energy/operation between the proposed 12-input ACN and CCN for 16 test vectors having different idealized capacitive loads. The table also demonstrates the percentage energy saving of the proposed ACN design compared to the CCN.

| Test vectors | Capacitive load $C_L$ (fF) | ACN (fJ) | CCN (fJ) | Savings (%) |
|---|---|---|---|---|
| TV1 | 426.7 | 127.2 | 1,439.1 | 91.2 |
| TV2 | 864.2 | 151.4 | 3,006.7 | 94.9 |
| TV3 | 505.1 | 130.7 | 1,498.2 | 91.3 |
| TV4 | 961.0 | 188.5 | 3,456.1 | 94.2 |
| TV5 | 186.3 | 95.1 | 365.3 | 73.9 |
| TV6 | 858.0 | 130.0 | 2,805.2 | 95.4 |
| TV7 | 935.9 | 154.4 | 3,109.3 | 95.0 |
| TV8 | 88.8 | 92.6 | 341.2 | 72.9 |
| TV9 | 298.8 | 116.0 | 769.7 | 84.9 |
| TV10 | 457.5 | 114.4 | 1,308.1 | 91.3 |
| TV11 | 943.0 | 159.7 | 3,143.4 | 94.9 |
| TV12 | 825.2 | 137.9 | 2,693.6 | 94.9 |
| TV13 | 838.0 | 143.9 | 2,714.3 | 94.7 |
| TV14 | 540.6 | 119.5 | 1,605.7 | 92.6 |
| TV15 | 344.9 | 118.5 | 905.4 | 86.9 |
| TV16 | 526.3 | 111.7 | 1,588.5 | 92.9 |

The two hardware-generated ACN membrane voltages (proposed and conventional) are in accord with one another, having a few $mV$ differences in the two $v_m^{\pm}$. As TL should not affect $v_m^{\pm}$, these differences are attributed to measurement variations arising from the $v_m^{\pm}$ peak voltage sampling time. Both hardware designs have an average absolute error of $14mV$ between the measured membrane voltages and that predicted by the theoretical model, with a maximum of $43mV$ for the proposed TL and $44mV$ for the conventional TL.

The test vectors (TV) (see Supplementary Table S1) are selected to test the system in a wide range of $v_{md}$ derived from different input patterns, to identify the loss of functionality and the minimum to maximum energy dissipation. The membrane voltages and the outputs for the 16 TV's are shown in Table 4. The conventional TL uses a combinational NOR-based SR latch which has an asymmetric gate switching threshold for low-to-high and high-to-low transitions. The NOR SR latch introduces an undefined switching threshold, causing the final output to latch at a different time. The final SR logic decision no longer corresponds strictly to the comparator threshold, degrading analytical accuracy and predictability. This can be seen in Table 1, which corresponds to a higher offset compared to Table 2. As a result, shown in Table 4, the ACN design with a conventional TL has a much larger $v_{md}$ between $16.8mV$ to $48.1mV$, resulting in a high number of functionality failures. In contrast, the proposed TL implementation employs a clocked SR latch that enforces a well-defined dynamic comparison window and preserves decision fidelity through controlled regeneration. Thus, improving the offset across process and temperature corners, which is shown in Tables 1, 2. Moreover, by synchronizing the comparison phase with clock-controlled input transistors, the proposed design suppresses premature regeneration and reduces the dominance of device mismatch during near-threshold operation. This results in a substantially lower rising offset, enabling reliable logic-1 generation for differential input voltages as small as $v_{md} \geq 6.3mV$.

## 4.2 Energy consumption

The total energy dissipation of an ACN consists of three components: 1) energy broadly due to the finite on-resistance of the bypass nMOS switch in the PCG, $E_{PCG}$ - see Supplementary Figure S7; 2) losses due to adiabatic charging and discharging of the synapse capacitance, $E_{AL}$ Younis (1994); 3) TL consumption, $E_{TL}$. The total energy dissipation $E_T$ in one cycle (excluding subthreshold leakage Chanda et al. (2015)) is given by:

$$E_T = E_{PCG} + E_{TL} + E_{AL} \tag{7}$$

$$E_{PCG} = \frac{1}{2} C_{PC} V_x^2 \left(1 - e^{2\frac{-t_{ON}}{R_{PC} C_{PC}}}\right) \tag{8}$$

$$E_{TL} = C_{TL} V_{DD}^2 \tag{9}$$

$$E_{AL} = C_L V_{DD}^2 \cdot \frac{\pi^2}{8} \cdot \frac{R_{syn} C_L}{T_r} \tag{10}$$

The first term in Equation 7 denotes the energy consumed in the PCG when the internal bypass switch is activated. This lost energy, $E_{PCG}$, is given by Equation 8, where $C_{PC}$ is the capacitance on the PC node, $R_{PC}$ is the ON resistance of the nMOS switch in PCG, $V_x$ is the residual voltage when the switch is on and, finally, $t_{ON}$ is the time when the PCG is in reset mode. The complete proof for the energy dissipation due to the sinusoidal ramp generator can be found in the foundation work of (Younis, 1994). The second term of $E_T$ is the TL energy loss, $E_{TL}$, due to the comparator and the latch, and is given by Equation 9. The comparator will switch state at every clock cycle, while the latch switches only once per operation. This gives almost a constant energy similar to standard inverter energy dissipation per switching event. Here, $C_{TL}$ is the equivalent total capacitance on the output TL node. The third term, $E_{AL}$, is an adiabatic loss for an equivalent RC network circuit under sinusoidal stimulation (Younis, 1994), which is given by Equation 10. Here, $C_L$ is the total capacitive load on the PC node due to the synapse. The resistance of the synapse switch is represented by $R_{syn}$. If the input is '$1'$', then $R_{syn}$ is a small value; otherwise, it will be very large, thus preventing the propagation of the PC signal to $v_m$. $T_r$ is the third parameter that defines the ramping time of the PC. The slower the system, the greater the energy efficiency. However, at some point, the leakage energy will start dominating (Teichmann, 2012).

In Equation 10, the $E_{AL}$ energy is shown to be proportional to $C_L^2$. Given that $C_L$ will be a function of the input pattern, and therefore $C_{on}$, the maximum energy dissipation of $E_{AL}$ is likely to occur near $max(C_L)$. In Supplementary Section 4, it is shown that, assuming an SPDT switch with negligible resistance, $C_L$ on each capacitive tree is given by the $C_{on}$ quadratic

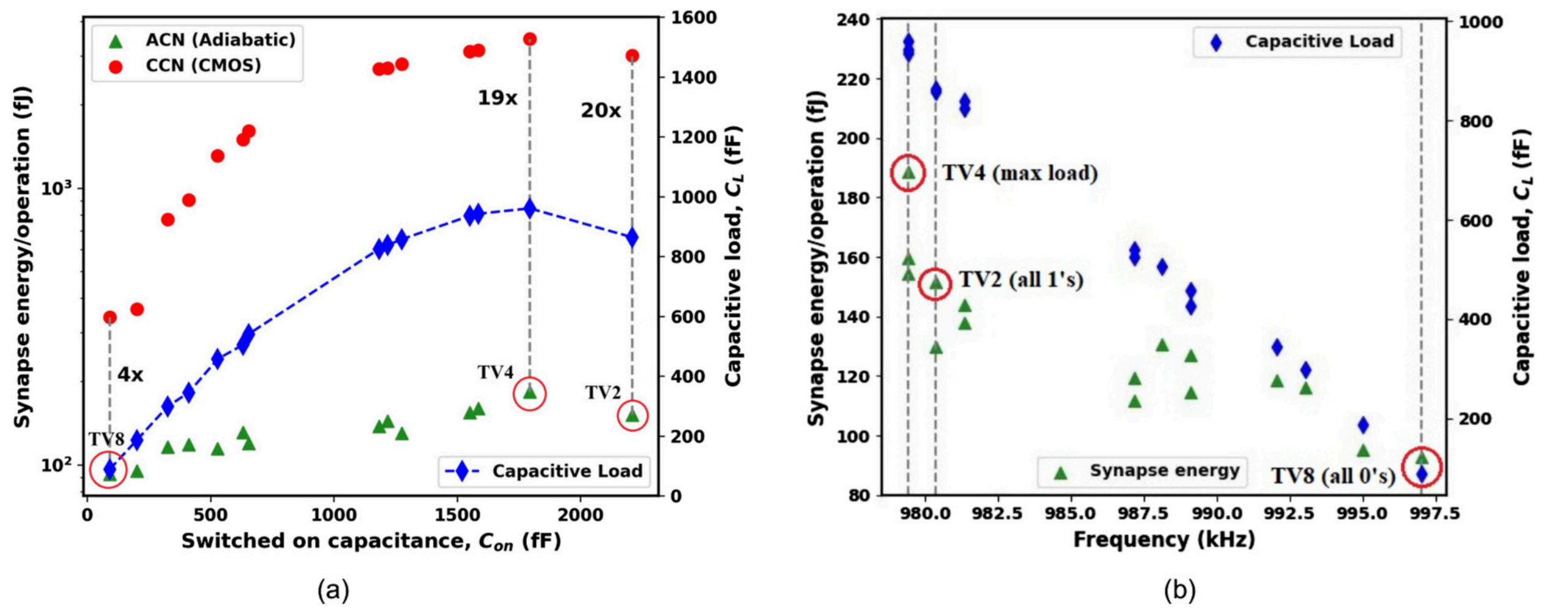

FIGURE 5
**(a)** The figure presents three plots, shown in red, green, and blue. The red and green curves correspond to the post-layout synapse energy per operation of the 12-input CCN and ACN, respectively, evaluated across multiple test vectors from the training set as a function of the effective switch-on capacitance. The blue curve illustrates the corresponding effective capacitive load, $C_L$, as a function of the switch-on capacitance for the same set of test vectors. Three representative test vectors are explicitly annotated in the figure: TV8, corresponding to the minimum switch-on capacitance $min(C_{on})$; TV2, corresponding to the maximum switch-on capacitance $max(C_{on})$; and TV4, which exhibits the maximum effective load capacitance $max(C_L)$. Each test vector corresponds to a unique capacitive load. **(b)** This figure presents two plots, shown in green and blue, showing the impact of the multiple test vectors from the training set on the frequency of operation. The nominal frequency was taken as $1MHz$. The green symbols correspond to the synapse energy per operation of the 12-input ACN, variations in frequency due to the change in the capacitive load for multiple test vectors.

TABLE 6 **Power clock parameters at different frequencies at maximum loading (TV4: worst case synapse energy). The channel width of the nMOS ($M_{PC}$) and $C_E$ in the PCG are constant and set to $10\mu m$ and $25pF$ respectively.**

| Nominal | Operating | PCG parameters | |
|---|---|---|---|
| Frequency ($MHz$) | Frequency ($MHz$) | $t_{ON}$ ($ns$) | $L_{PC}$ ($mH$) |
| 0.10 | 0.0986 | 600.0 | 100.0 |
| 0.50 | 0.4902 | 120.0 | 4.0 |
| 1.0 | 0.9794 | 60.0 | 1.0 |
| 10.0 | 9.8100 | 6.0 | 0.010 |
| 100.0 | 98.0400 | 0.6 | 0.0001 |

$$C_L^{\pm} = -\frac{C_{on}^{\pm} C_{on}^{\pm}}{C_A^{\pm}} + C_{on}^{\pm} \quad (11)$$

which holds for $C_L^{\pm} > 0$ as $C_A^{\pm} \geq C_{on}^{\pm}$ and gives $max(C_L)$ when $C_{on}^{\pm} = C_A^{\pm}/2$.

The total synapse energy per operation measured in post-layout simulation for the ACN and CCN implementations is provided in Table 5. The reported synapse energy per operation includes the contribution of the PCG. The TL energy is constant for both implementations and is thus not included in the results. The capacitive load on the PC are computed using Equation 11. The derivation of the capacitive loading of Equation 11 is discussed in Section 4 of the Supplementary document. The ACN network demonstrates average energy savings of more than 90% typically compared to the CCN. The data points corresponding to TV8 and TV5 deviate from this trend and can be considered outliers, likely arising from input capacitance loading conditions that differ from the nominal regime. However, the ACN energy reported in Table 5 does not include the CMOS inverter energy shown in Figure 2b. This non-adiabatic inverter circuit (powered by a DC voltage) would cause the total synapse energy dissipation to increase by ≈ 30–35%. It has been excluded here, as in multilayer networks, the complementary output from the previous layers' threshold logic can be used directly instead. However, in the case of multilayer CCN, extra circuitry between layers is mandatory to provide transition (either zero to $V_{DD}$ or vice-versa), enabling the synapse capacitors to compute the membrane voltages. The description of the CCN design is presented in Supplementary Section 2 and the circuit diagram is shown in Supplementary Figure S6.

The capacitive load strongly influences both energy dissipation and recovery in an adiabatic system. While increasing capacitive load raises the total stored energy, it also increases the amount of charge returned to the PC, with dissipation remaining limited by the PC ramp time, $T_r$ as described by the adiabatic loss energy expression in Equation 10. For TV8, where all inputs are zero, the effective load capacitance $C_L$ is dominated by the bias capacitors ($C_b^{\pm}$), resulting in limited charge recovery, while the energy dissipation is primarily set by the PCG. Consequently, for test vectors such as TV8 and TV5 with minimal active synapses, only marginal energy savings are achieved which is still very substantial, underscoring the importance of adiabatic charge recovery for energy efficiency.

The plot in Figure 5a represents the data from Table 5 with respect to the on-capacitance $C_{on}$. As predicted, the maximum measured synapse energy is around $max(C_L)$, rather than all 1's ($TV2$) input pattern. This is different from ACAN (Maheshwari et al., 2021b; Maheshwari et al., 2022) where $C_L$ will increase monotonically with $C_{on}$. We further note that despite a nominal frequency of $1MHz$, differences in the PC capacitive load induced by different test vectors mean that the actual frequency changes slightly for each row of Table 5. The Figure 5b reflects the relationship between operational frequency, capacitive load and synapse energy. The actual frequency ranges from $979kHz$ (TV4: maximum load) to $997kHz$ (TV8: all zero inputs), which is an approximately 2% variation off the nominal frequency.

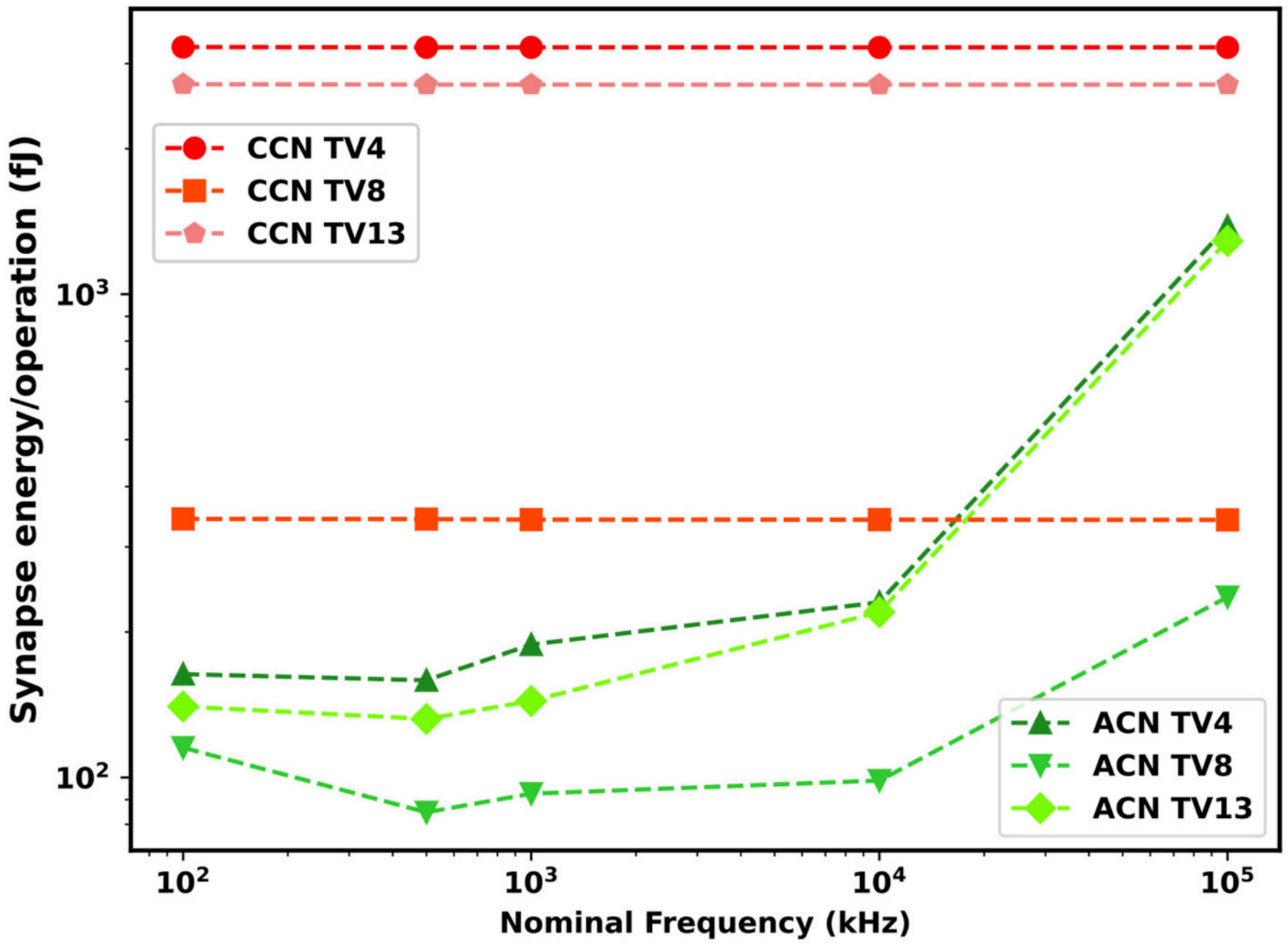

FIGURE 6
Total synapse energy/operation versus operating frequency across 3 TVs for 12-input ACN and CCN. The reddish colour shades represent the CCN energy plots, while the greening shades represent ACN energy. The two test vectors are chosen corresponding to the minimum switch-on capacitance $min(C_{on})$ TV8, and maximum load capacitance $max(C_L)$, TV4. since TV4 and TV8 has final output zero, the third test vector was chosen randomly to be TV13 that has a final output 1 and the total capacitive load value close to TV2.

## 4.3 Frequency scaling

This subsection presents the results for a broader range of scenarios, where the nominal frequency varies from our baseline scenario of 1 $MHz$. The added frequencies are: $100kHz$, $500kHz$, $10MHz$ and $100MHz$. For all frequencies, some coarse-grain optimization is done in the balance between the capacitance and inductance of the PCG and is reported in Table 6. Due to the synapse loading on the PC, the actual operating frequency across the range drops by about 2% vs. nominal (see Figure 5b). The total 12 1-bit input synapse energy per operation for ACN and CCN across operating frequency for 3 test vectors is reported in Figure 6. It can be clearly seen that with decreasing frequency (i.e., increasing ramping time ($T_r$) and period ($t_{ON}$)), energy dissipation reduces. It is observed that at $100kHz$, the system energy starts to increase, which mainly arises due to the PCG non-idealities (Raghav et al., 2025). We see ~ $7.67 fJ/MHz$ average change in energy for TV4 for the frequency range [0.5, 10]$MHz$, which shrinks to ~$4.76 fJ/MHz$ within [1, 10]$MHz$. Significant energy savings of > 90% are clear within [0.5, 10]$MHz$.

## 4.4 Voltage scaling

In both adiabatic and non-adiabatic logic, the energy dissipation is directly proportional to the square of the power supply. Thus, a further energy reduction can be achieved if the supply voltage is reduced. In adiabatic logic, voltage scaling affects both non-adiabatic energy dissipation in the PCG, $E_{PCG}$, and the adiabatic loss, $E_{AL}$. In an adiabatic system, energy consumption can be reduced under voltage scaling by adjusting key parameters: lowering the ON resistance of the synapse transistor ($R_{syn}$), increasing the resistance in the PCG ($R_{PC}$)—achieved by reducing the width of the nMOS switch in the PCG—and decreasing the supply voltage ($V_{DD}$), which linearly influences the node voltage $V_x$. The on-resistances, $R_{syn}$ and $R_{PC}$ are inversely proportional to $(V_{GS} - V_{th})$. As the supply voltage decreases, the $E_{AL}$ tends to increase while $E_{PCG}$ tends to decrease. However, the increase in $E_{AL}$ is balanced by the square of the supply voltage and the capacitive load. On the other hand, on decreasing $V_{DD}$, $V_x$ also decreases, thus the overall energy dissipation of the PCG decreases.

The impact of supply voltage scaling on synapse energy dissipation is shown in Figure 7. The main trends are that: a) TV8 (all-off) follows a relatively smooth drop in energy, similar to what we see in the CCN and b) TV4 and 13 (actively loaded system) show that overall voltage downscaling does lead to a drop in energy, with levels below 1.3 V showing faster drops than the CCN.

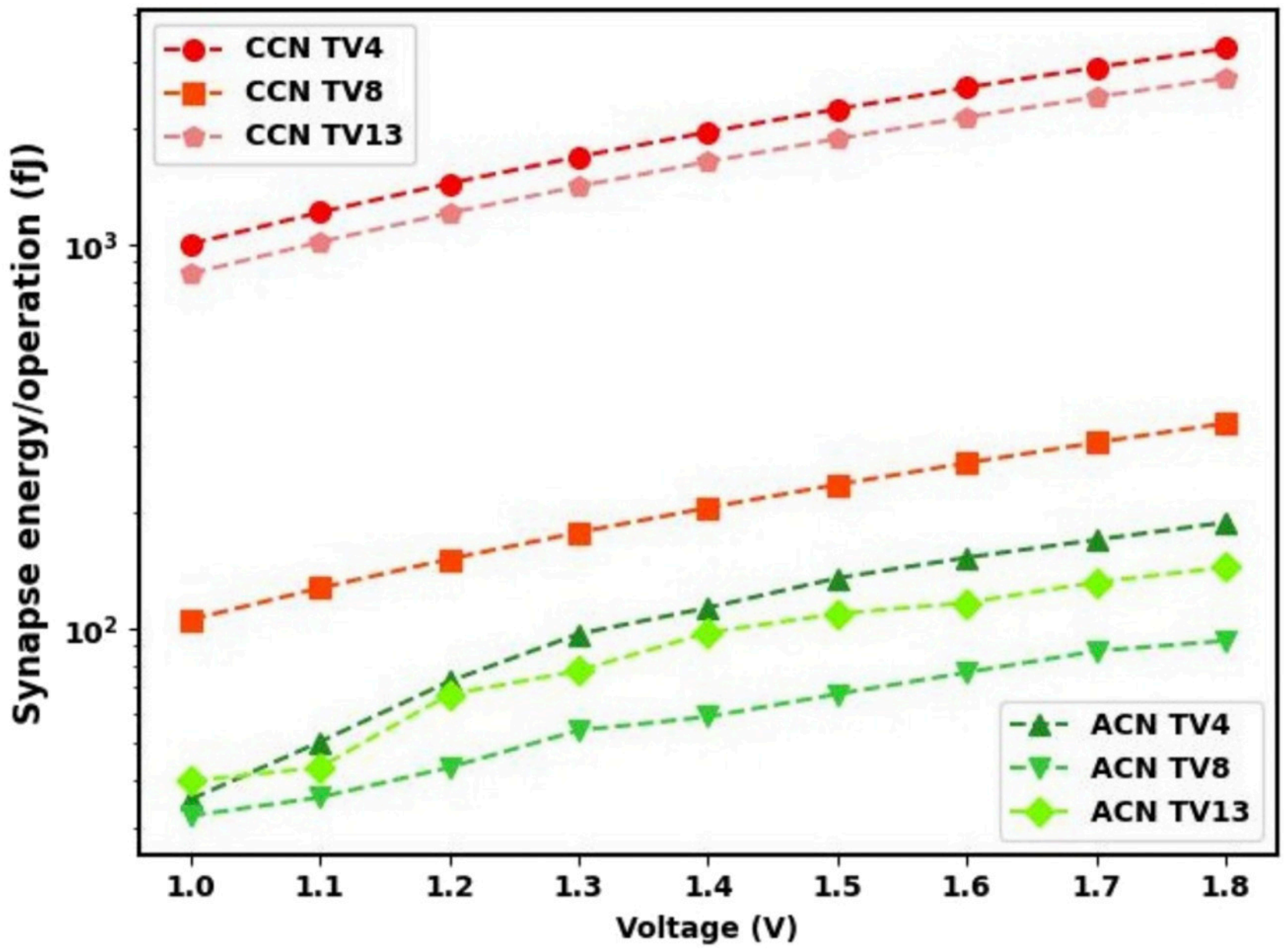

FIGURE 7
The figure illustrates the total synapse energy per operation as a function of supply voltage, scaled from 1.8$V$ down to 1.0$V$, for three representative test vectors, TV4, TV8 and TV13. Results are shown for both 12-input CCN and the ACN, with the red dashed curves corresponding to the CCN and the green dashed curves corresponding to the ACN. The comparison highlights the impact of voltage scaling on energy consumption and emphasizes the relative energy-efficiency advantage of the ACN across the examined operating range.

At 1$V$, the energy of TV4 and TV13 starts to approach that of TV8, which indicates that $E_{AL}$ has almost decreased to zero. Supplementary Table S2 shows ACN energy savings vs. CCN under voltage scaling. The adiabatic circuit shows an average saving of ~ 95% for all the test vectors, except for TV8 (all 0's). As no switching occurs in the synapse circuitry with TV8, the dissipated energy is only due to $E_{PCG}$.

## 4.5 Statistical analysis

Standard Monte Carlo simulations consisting of 1,000 runs for the proposed ACN and CCN post-layout designs were tested based on the worst-case loading/energy input vector pattern, TV4. We considered the global process variations (wafer-to-wafer and run-to-run) and mismatch (non-uniformity across individual wafers) Fischer et al. (2005). The statistical analysis uses the Low Discrepancy Sequence (LDS) sampling method provided in the Cadence tool, which has a uniform sample space coverage. Figure 8a clearly shows a right-skewed energy distribution for the ACN. The distribution appears to have a long tail, meaning a large number of occurrences are far from the mean value. This implies that, under certain conditions, an unexpectedly large synapse energy may be generated.

Figure 8b is a quantile-quantile (or Q-Q) plot to verify that the ACN synapse energy Monte Carlo results in Figure 8a are indeed not normally distributed. A correlation coefficient of 0.89 and a very small p-value reject the hypothesis that the data is normally distributed. The same analysis is also carried out for the CCN design. The distribution in Figure 8c appears to be normally distributed, which is confirmed by its Q-Q plot shown in Figure 8d, where all the data points lie on the line. A correlation coefficient of 1 indicates linearity and confirms the data as normally distributed. The coefficient of variation (CV) is defined as the ratio of the standard deviation to the mean. As the data points are well spread out, the CV for the ACN is calculated as 12.55, whereas for CCN it is 4.85. It is inferred that the skewed energy distribution in an adiabatic system originates from the conditional and nonlinear nature of charge recovery, where small variations in device parameters, such as threshold voltage, or timing can asymmetrically degrade recovery efficiency. Because energy dissipation (see Equation 10) depends multiplicatively on effective resistance, $R_{syn}$, capacitance, $C_L$, and ramping time, $T_r$, rare recovery failures produce high-energy outliers, resulting in an intrinsically skewed distribution.

# 5 Conclusion

In this paper, we have introduced a novel, differential, adiabatic, switched-capacitor artificial neuron combined with a new threshold logic design. In comparison to previous work, such as ACAN, new functionality has been added in the form of support for negative-valued ANN weights through the differential DTSC ACN architecture. This could potentially mean fewer neurons are required overall to implement an ANN to the same level of functional performance. This ACN design also introduced new functionality in the form of a two-stage TL latch design to implement a binary activation function.

In terms of accuracy, the paper has shown that weights from a real ANN model can be easily mapped to synapse capacitance values to perform the same operation. The post-layout simulations have provided a good correlation between a theoretical software model and post-layout results. The new TL design was also shown to reduce errors in the ACN output due to its reduced offset compared to the conventional TL design. The differential design introduces some additional robustness that was missing in the previous ACAN design. The ACAN circuit Maheshwari et al. (2022) was susceptible to errors if the PC voltage changed for any reason. This is because the ACAN uses a fixed absolute DC threshold voltage, which the TL compares against its single membrane potential. The ACN eliminates this limitation by using a differential tree topology, with any errors with $V_{pc}(t)$ affecting both trees in the same way and consequently not affecting the output. It also means that the ACAN non-zero DC threshold voltage does not need to be supplied to each neuron. Furthermore, we have explored some important design parameters of this new ACN design, such as PC voltage and nominal frequency that can be adjusted without altering the mapped capacitor configuration. Although the new TL design in this work shows robustness across process-voltage-temperature corners, its behavior can be influenced by input-referred offset voltage, noise, and finite gain, which introduces uncertainty around the decision boundary. Additionally, delay and metastability become critical in the TL comparator when operating at high speeds or with slowly varying input signals. The non-idealities, related to kickback noise and loading, constrain the achievable precision and reliability of threshold-based decision making in practical implementations. A comprehensive characterization and optimization of these effects/impact represent important and timely avenues for future investigation, and are expected to provide valuable insight for advancing high-precision and scalable threshold-logic implementations.

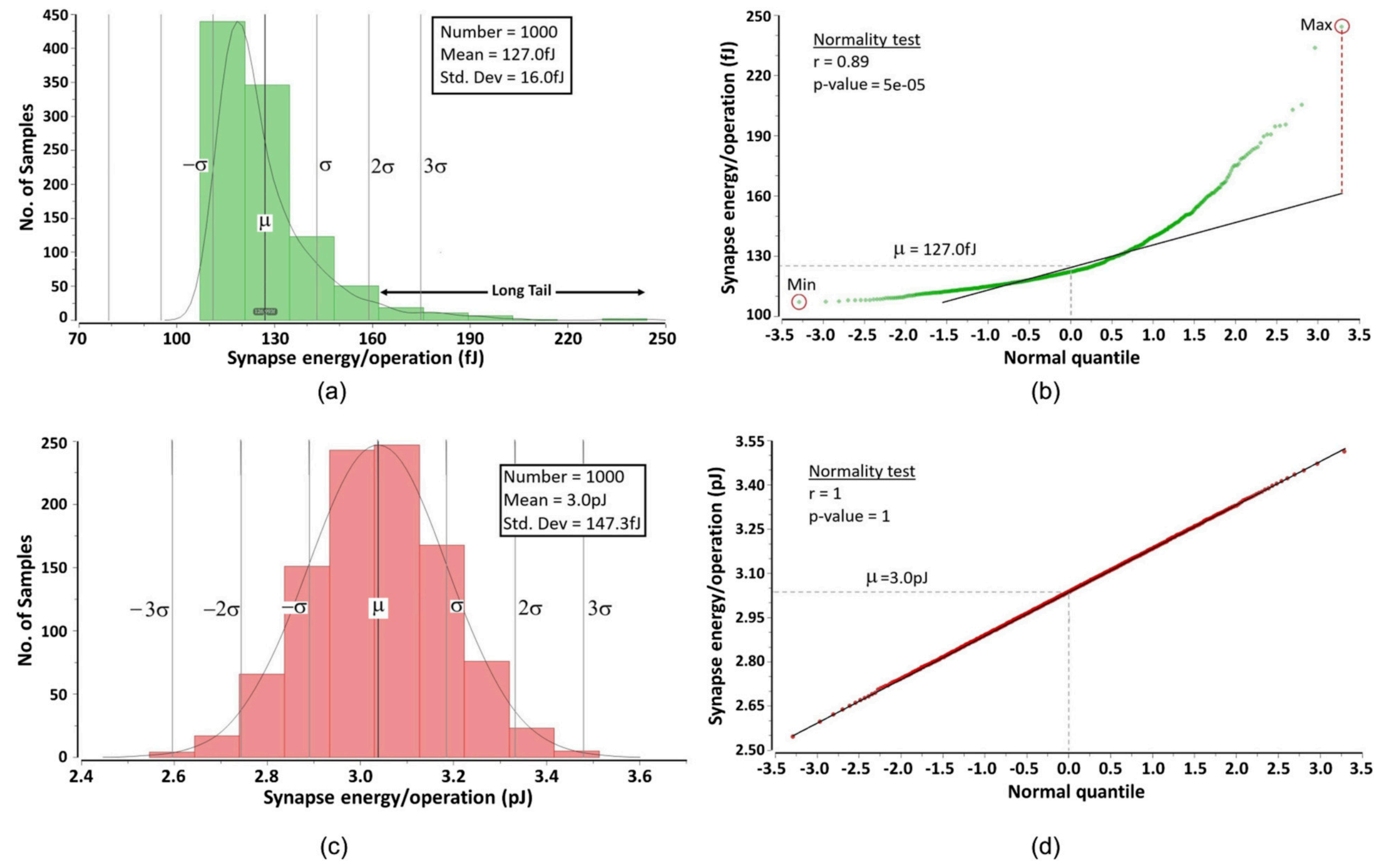

FIGURE 8
Worst-case input vector, TV4 synapse energy distribution over 1,000 runs with mean, $\mu$, and standard deviation, $\sigma$ and their Q-Q plots for ACN and CCN designs. **(a)** Right-skewed 12-input ACN synapse energy distribution. Some of the data points lie outside $+3\sigma$. **(b)** Q-Q plot compares observed synapse energy data (vertical axis) to a statistical normal quantiles theoretical data (horizontal axis). The deviation from the straight line indicates skewness, demonstrating that the energy data is not distributed as a standard normal. **(c)** Synapse energy/operation appears normally distributed for the CCN. **(d)** Q-Q plot compares observed synapse energy data (vertical axis) to a statistical normal quantiles theoretical energy data (horizontal axis). The linearity of the points with the straight line suggests that the data is normally distributed.

In terms of energy efficiency, the post-layout analysis of a single neuron with 12 input synapses shows significant savings between 90% and 95% compared to an equivalent non-adiabatic CCN implementation. Monte Carlo simulations over 1,000 samples resulted in a skewed synapse energy distribution for the ACN. However, the CCN design showed normally distributed synapse energies. The non-normal distribution of the ACN is a result of its energy dependency on the synapse switch resistance and ramping time/frequency. However, it should be noted that even the ACN samples in the long tail of the distribution still outperformed those from the CCN.

Overall, the ACN represents a significant step forward in practical, high-energy-efficient, and accurate ANN computation. This paper has focused on a fixed capacitor array implementation, but with the recent advancement in mem-impedance devices, memcapacitors seem to be an interesting choice for SC networks due to their tunable properties, and have already been deployed with parallel multiply-accumulate operations (Demasius et al., 2021), in integrate-and-fire neural networks (Pershin and Ventra, 2014) and AN synapse neuro-transistors (Wang et al., 2018).

## Data availability statement

The original contributions presented in the study are included in the article/Supplementary Material, further inquiries can be directed to the corresponding author.

## Author contributions

SM: Conceptualization, Formal Analysis, Investigation, Methodology, Validation, Writing – original draft, Writing – review and editing. MS: Conceptualization, Data curation, Formal Analysis, Software, Validation, Writing – original draft, Writing – review and editing. HR: Investigation, Methodology, Writing – original draft, Writing – review and editing. TP: Writing – review and editing. AS: Funding acquisition, Project administration, Supervision, Writing – review and editing.

## Funding

The author(s) declared that financial support was received for this work and/or its publication. This work has been, in part, funded by Defence Science and Technology Laboratory (Dstl), United Kingdom.

## Conflict of interest

The author(s) declared that this work was conducted in the absence of any commercial or financial relationships that could be construed as a potential conflict of interest.

## Generative AI statement

The author(s) declared that generative AI was not used in the creation of this manuscript.

Any alternative text (alt text) provided alongside figures in this article has been generated by Frontiers with the support of artificial intelligence and reasonable efforts have been made to ensure accuracy, including review by the authors wherever possible. If you identify any issues, please contact us.

## Publisher's note

## Supplementary material

The Supplementary Material for this article can be found online at: https://www.frontiersin.org/articles/10.3389/felec.2026.1743265/full#supplementary-material